\documentclass[a4paper,11pt]{article}
\usepackage{pos}

\title{Non-perturbative Collins-Soper kernel: Chiral quarks and Coulomb-gauge-fixed quasi-TMD}
\ShortTitle{Non-perturbative CS kernel: Chiral quarks and Coulomb-gauge-fixed quasi-TMD}

\author[a]{Swagato Mukherjee}
\author[b]{Dennis Bollweg}
\author[c]{Xiang Gao}
\author[c]{Yong Zhao}

\affiliation[a]{Physics Department, Brookhaven National Laboratory, Upton, 11973 NY, USA}

\affiliation[b]{Computational Science Initiative, Brookhaven National Laboratory, Upton, 11973 NY, USA}

\affiliation[c]{Physics Division, Argonne National Laboratory,  Lemont, IL 60439, USA}
\emailAdd{dbollweg@bnl.gov}
\emailAdd{gaox@anl.gov}

\abstract{We present the first lattice QCD calculation of the rapidity anomalous dimension of transverse-momentum-dependent distributions (TMDs), 
i.e. the Collins-Soper (CS) kernel, employing the recently proposed Coulomb-gauge-fixed quasi-TMD formalism as well as a chiral-symmetry preserving lattice discretization.

This unitary lattice calculation is conducted using the domain wall fermion discretization scheme, a fine lattice spacing of approximately 0.08 fm,
and physical values for light and strange quark masses. The CS kernel is determined analyzing the ratios of pion quasi-TMD wave functions (quasi-TMDWFs)
at next-to-leading logarithmic (NLL) perturbative accuracy.

Thanks to the absence of Wilson-lines, the Coulomb-gauge-fixed quasi-TMDWF demonstrates a remarkably slower decay of signals with increasing quark separations.
This allows us to access the non-perturbative CS kernel up to transverse separations of 1 fm. For small transverse separations, our results agree well with perturbative predictions. 
At larger transverse separations, our non-perturbative CS kernel clearly favors certain global fits.}

\FullConference{31st International Workshop on Deep Inelastic Scattering (DIS2024)\\
 8–12 April 2024\\
Grenoble, France\\}

\begin{document}
\maketitle

\section{Introduction}
The Collins-Soper kernel, describing the rapidity scale evolution of transverse-momentum-dependent distributions (TMDs), is essential in
interpreting experimental data of TMD measurements across different energy scales and connecting them to theory predictions. 
While the CS kernel can be extracted through global analysis of semi-inclusive deep inelastic scattering and Drell-Yan processes, parameterizing
these often introduces model dependencies, particularly at low transverse momenta. Therefore, model-independent, first-principle calculations 
via lattice Quantum Chromodynamics are highly desirable. 

Light-cone correlations, such as TMDs, can be accessed in lattice QCD through the framework of Large-Momentum Effective Theory (LaMET) \cite{Ji:2013dva,Ji:2014gla}.
TMDs defined on the light-cone are related to quasi-TMDs in the large momentum and $\eta\rightarrow \infty$ limit through perturbative factorization \cite{Ji:2018hvs,Ebert:2018gzl}. The quasi-TMDs involve matrix elements of equal-time gauge-invariant operators
\begin{align}
  \label{eq:GIMAT}
    \mathcal{O}_{\Gamma}^{GI}(\mathbf{b};\eta)=\bar{\psi}\left(\frac{\mathbf{b}}{2}\right)\Gamma W_{\sqsupset }\left(\frac{\mathbf{b}}{2},-\frac{\mathbf{b}}{2},\eta\right)\psi\left(-\frac{\mathbf{b}}{2}\right),
\end{align}
with $\mathbf{b}=(b_{\perp},b_z)$ and $W_{\sqsupset}$ a staple-shaped Wilson-line of length $2\eta+b_z$. 

Despite many advancements, lattice QCD calculations of TMDs remain challenging. A large momentum $P_z$ is required to suppress power corrections, which requires fine lattice spacings and incurs large numerical costs. 
Furthermore, the space-like Wilson-line introduces a signal-to-noise ratio that decays exponentially with the total length of the Wilson-line, 
making it particularly difficult to complement phenomenological analyses in the large $b_{\perp}$ region. Complicated behavior
of the Wilson-line under renormalization only exacerbates these problems.

Recently, a new approach for Parton physics calculations based on Coulomb-gauge-fixed operators has been proposed \cite{Gao:2023lny,Zhao:2023ptv} which eliminates the 
need for Wilson-lines and thereby simplifying lattice calculations significantly. In the Coulomb-gauge-fixed approach, quasi-TMDs defined via matrix elements of the form 
\begin{align}
  \label{eq:CGMAT}
    \mathcal{O}_{\Gamma}^{CG}(\mathbf{b})=\bar{\psi}\left(\frac{\mathbf{b}}{2}\right)\Gamma \psi\left.\left(-\frac{\mathbf{b}}{2}\right)\right\vert_{\nabla\cdot \mathbf{A}=0},
\end{align}
subject to the Coulomb gauge condition but without the presence of a Wilson-line, are studied. It has been shown that these quasi-TMDs can be related to the light-cone TMDs in physical gauge \cite{Zhao:2023ptv}. 

The Collins-Soper kernel does not depend on the external states and can therefore be extracted, for example, via the pion quasi-TMD wavefunction, defined as the Fourier transform of 
\begin{align}
  \tilde{\Phi}_{\Gamma}(b_{\perp},b_z,P_z,\mu)=\langle \Omega\vert \mathcal{O}_{\Gamma}(\mathbf{b})\vert\pi^{+};P_z\rangle,
\end{align}
with momentum $\mathbf{P}=(0,0,P_z)$. 
The pion quasi-TMDWF can be related to the light-cone TMDWF $\phi(x,b_{\perp},\zeta,\mu)$ via 
\begin{align}
  \frac{\tilde{\phi}_{\Gamma}(x,b_{\perp},P_z,\mu)}{\sqrt{S_r(b_{\perp},\mu)}}&=H(x,\bar{x},P_z,\mu)\phi(x,b_{\perp},\zeta,\mu)\exp\left[\frac{1}{4}\left(\ln\frac{(2xP_z)^2}{\zeta}+\ln\frac{(2\bar{x}P_z)^2}{\zeta}\right)\gamma^{\overline{\mathrm{MS}}}(b_{\perp},\mu)\right]\\\nonumber
  &+\mathcal{O}\left(\frac{\Lambda_{QCD}^{2}}{(xP_z)^2},\frac{1}{(b_{\perp}(xP_z))^2},\frac{\Lambda_{QCD}^2}{(\bar{x}P_z)^2},\frac{1}{(b_{\perp}(\bar{x}P_z))^2}\right),
\end{align}
where $H(x,\bar{x},P_z,\mu)$ is a matching kernel that can be computed from perturbation theory \cite{Zhao:2023ptv}, $S_r(b_{\perp},\mu)$ is the reduced soft function and $\gamma^{\overline{\mathrm{MS}}}(b_{\perp},\mu)$ is the Collins-Soper kernel. 

We will use this relation to extract the Collins-Soper kernel through the ratios of quasi-TMDWFs at different momenta in the next sections.
\section{Quasi-TMDWF}
We compute the bare matrix elements of the pion quasi-TMDWF via the two-point correlation functions 
\begin{align}
  &C_{\pi\mathcal{O}}(t_s,b_{\perp},b_z,P_z)=\langle \mathcal{O}_{\Gamma}(\mathbf{b},\mathbf{P},t_s)\pi^{\dagger}(\mathbf{y}_0,0)\rangle,\\
  &C_{\pi\pi}(t_s,P_z)=\langle \pi(\mathbf{P},t_s)\pi^{\dagger}(\mathbf{y}_0,0)\rangle,
\end{align}
which we compute with both quasi-TMD operator cases \eqref{eq:GIMAT} and \eqref{eq:CGMAT}. $\mathbf{y}_0$ denotes the source position, $t_s$ is the time separation and we chose 
$\Gamma=\gamma_t\gamma_5$ because it does not suffer from operator mixings due to chiral symmetry breaking. 
We utilize the 64I gauge field ensemble generated by the RBC and UKQCD collaborations which uses 2+1-flavor Domain wall fermions on a $N_s^3\times N_t \times L_s= 64^3\times 128\times 12$ lattice with physical quark masses and a lattice spacing of $a=0.0836$ fm.
In total, 64 gauge field configurations were analyzed. In order to obtain the gauge invariant matrix elements we use staple shaped Wilson-lines with $\eta=12a$ and employ Wilson flow with a flow time $t_F=1.0$ to enhance signal-to-noise ratio. For more details on the computational setup, we refer to \cite{Bollweg:2024zet}.

We form the ratios 
\begin{align}
  \label{eq:fitratios}
  R(t_s,b_{\perp},b_z,P_z)=\frac{-iC_{\pi\mathcal{O}}(t_s;b_{\perp},b_z,P_z)}{C_{\pi\pi}(t_s;P_z)},
\end{align}
which give access to the bare matrix element $\tilde{\phi}^{B}(b_{\perp},b_z,P_z,a)$ in the $t_s\rightarrow \infty$ limit. We fit the data with a two-state fit and
then form renormalization group invariant ratios 
\begin{align}
  \label{eq:RGratios}
  \tilde{\Phi}^{GI}(b_{\perp},b_z,P_z)=\frac{\tilde{\phi}^{B}_{GI}(b_{\perp},b_z,P_z,\eta,a)}{\tilde{\phi}^{B}_{GI}(b_{\perp},0,0,\eta,a)},\;\;\mathrm{and}\;\;  \tilde{\Phi}^{CG}(b_{\perp},b_z,P_z)=\frac{\tilde{\phi}^{B}_{CG}(b_{\perp},b_z,P_z,a)}{\tilde{\phi}^{B}_{CG}(b_{\perp},0,0,a)}.
\end{align}
In the Coulomb-gauge-fixed case, the bare matrix elements pick up an overall wave function renormalization factor that is independent of $b_{\perp}$ and $b_z$. In the gauge-invariant case, however, the Wilson-line requires the removal of additional pinch pole singularities, linear divergences and cusp divergences, which depend on the total length of the Wilson-line, $2\eta+b_{\perp}$ and are cancelled in equation \eqref{eq:RGratios}.

\begin{figure}[ht]
\centering
\includegraphics[width=.49\textwidth]{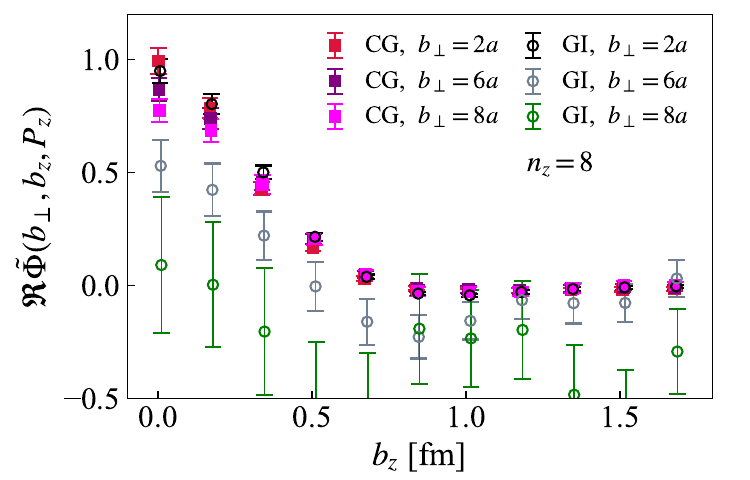}
\includegraphics[width=.49\textwidth]{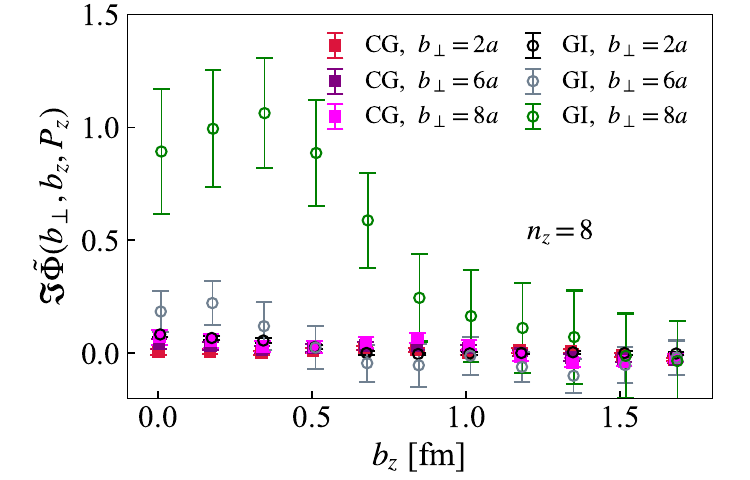}
\caption{Real (left) and imaginary (right) parts of the renormalized quasi-TMDWF matrix elements at $b_{\perp}=2a,6a,8a$ with momentum $n_z=8$ for the
Coulomb-gauge-fixed (filled squares) and gauge invariant case (open circles).}
\label{fig:matrixelements}
\end{figure}

We show the results for the renormalized matrix elements for the largest available momentum $P_z=1.85$ GeV and $b_{\perp}=2a,6a,8a$ as a function of $b_{z}$ in Figure \ref{fig:matrixelements}.
In the Coulomb-gauge-fixed case, we maintain a good signal up to the highest $b_{\perp}$ values, whereas the signal-to-noise ratio deteriorates quickly in the gauge-invariant case as seen most prominently in the $b_{\perp}=8a$ data points (green) in Figure \ref{fig:matrixelements}.
Furthermore, the imaginary part of the renormalized quasi-TMDWF matrix elements remains zero within errors in the Coulomb-gauge-fixed case while the gauge-invariant case shows non-zero values. 
Additionally, the real part of the matrix elements approach zero as $b_z$ is increased, a fact that we exploit to perform the Fourier transform to $x$-space after interpolating our quasi-TMDWF matrix elements with first-order splines.  
\begin{figure}[ht]
  \centering
  \includegraphics[width=.49\textwidth]{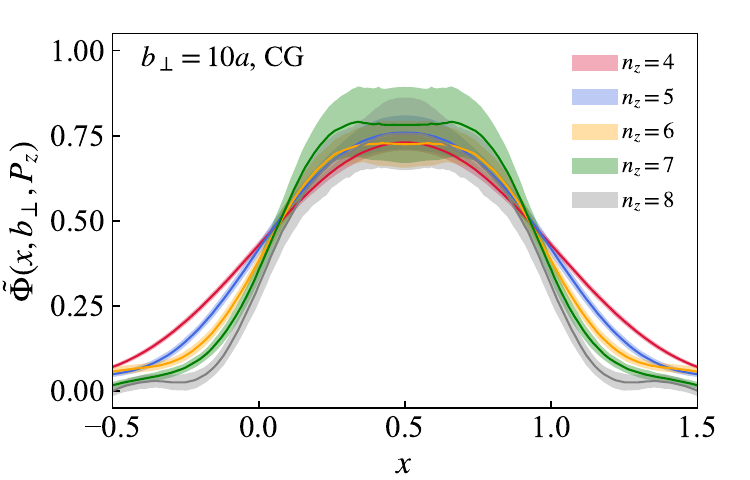}
  \includegraphics[width=.49\textwidth]{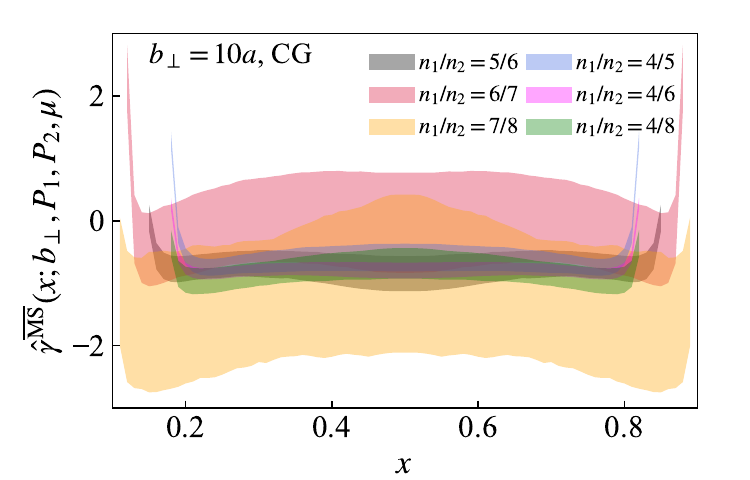}
  \caption{Left: $x$-space quasi-TMDWF matrix elements determined via the Coulomb-gauge-fixed method at momenta $n_z\in[4,8]$ and transverse separations $b_{\perp}=10a$. Right: Collins-Soper kernel estimators $\hat{\gamma}^{\overline{\mathrm{MS}}}(x,b_{\perp},P_1,P_2,\mu)$ derived from ratios of the quasi-TMDWFs.}
  \label{fig:xspaceQTMD}
\end{figure}

As seen in Figure \ref{fig:xspaceQTMD} (left), we find reasonable signal-to-noise ratios up to $b_{\perp}=10a$ although the noise does increase with increasing momentum. Furthermore, the quasi-TMDWFs shrink into the physical $x\in[0,1]$ region as the momentum is increased, consistent with the expectations from LaMET that the quasi-TMDWF approaches the light-cone TMDWF in the large momentum limit.
\section{The Collins-Soper kernel}
We define estimators for the CS kernel via a ratio of quasi-TMDWFs at two finite momenta $P_1$ and $P_2$,
\begin{align}
    \label{eq:estimator}
    \hat{\gamma}^{\overline{\mathrm{MS}}}(x,b_{\perp},P_1,P_2,\mu)=\frac{1}{\ln(P_2/P_1)}\ln\left[\frac{\tilde{\Phi}(x,b_{\perp},P_2)}{\tilde{\Phi}(x,b_{\perp},P_1)}\right]+\delta\gamma^{\overline{\mathrm{MS}}}(x,\mu,P_1,P_2).
\end{align}

We applied the perturbative corrections $\delta\gamma^{\overline{\mathrm{MS}}}$ derived from next-to-leading logarithm matching kernels for the Coulomb-gauge-fixed case as only one-loop non-cusp computations are available.
The $\overline{\mathrm{MS}}$-scale has been set to $\mu=2$ GeV. In Fig.\ref{fig:xspaceQTMD} (right), we show the estimators obtained from \eqref{eq:estimator} for the Coulomb-gauge-fixed case and various pairs of momenta $n_1$ and $n_2$ in $x$-space.
Robust plateaus are found in the moderate $x$-region even up to the largest $b_{\perp}$ used. The estimators seem to diverge in the end-point regions of small and large $x$, signaling a breakdown of factorization. The length of these plateaus, however, increases with increasing momentum. 

No momentum dependence is found in the case of large $b_{\perp}$, indicating well suppressed power corrections even in the presence of larger statistical uncertainties at large momentum. 

\begin{figure}[ht]
    \centering
    \includegraphics[width=.65\textwidth]{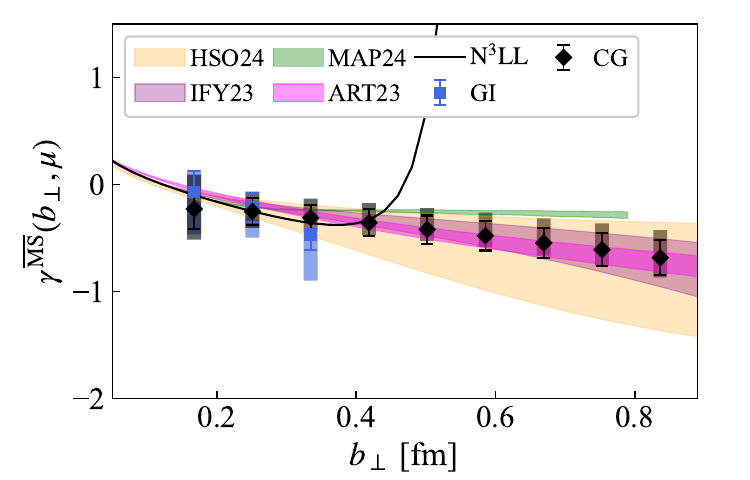}
    \caption{The CS kernel as a function of transverse separation $b_{\perp}$. Results from the CG quasi-TMDWFs
     are shown as black points and patches. The results from the GI quasi-TMDWFs are shown as blue points and patches. Phenomenological parameterizations of experimental
     data are shown in colored bands, including data from MAP24\cite{Bacchetta:2024qre}, ART23\cite{Moos:2023yfa}, IFY23\cite{Isaacson:2023iui} and HSO24\cite{Aslan:2024nqg}. Furthermore, perturbative results
     up to $N^3LL$ are shown.}
     \label{fig:CSresult}
\end{figure}

Finally, we average over the estimator $\hat{\gamma}^{\overline{\mathrm{MS}}}(x,b_{\perp},\mu,P_1,P_2)$ within $x\in [x_0,1-x_0]$ and varying $n_1$ and $n_2$ pairs, considering only pairs with $n_2-n_1 = 1$. $x_0$ is determined by requiring $2x_0P_zb_{\perp}>1$ and $2x_0P_z > 0.7 $ GeV. Averages over $x$ and different $n_1/n_2$ pairs are carried out for each bootstrap sample of gauge field configurations and the results are quoted from the median and 68\% confidence limit of the bootstrap sample distribution.
The full results for the Collins-Soper kernel determined with the Coulomb-gauge-fixed method are shown in Figure \ref{fig:CSresult} as black points and black patches, showing the results when excluding or including $n_1/n_2=6/7$ and $7/8$, respectively.
Results for the gauge-invariant approach are shown via blue points and patches and are consistent with the Coulomb-gauge-fixed method at small $b_{\perp}$, though results beyond $b_{\perp}=4a$ are excluded due to too large noise.

\section{Conclusion}
We performed the first lattice QCD calculation of the Collins-Soper kernel using the recently proposed Coulomb-gauge-fixed quasi-TMD method with Domain-Wall fermions and physical quark masses.
We find the Coulomb-gauge-fixed method to produce significantly better signal-to-noise ratios compared to the gauge-invariant method, allowing us to extend our calculations of the Collins-Soper kernel to large values of $b_{\perp}$. 
Additionally, our results agree well with the established gauge-invariant method as well as recent phenomenological parameterizations of experimental data and we are able to reproduce the linear $b_{\perp}$-dependence found in the phenomenological parameterizations.

\section{Acknowledgements}
This material is based upon work supported by The U.S. Department of Energy, Office of Science, Office of Nuclear Physics through Contract No. DE-SC0012704, 
Contract No. DE-AC02-06CH11357, and within the frameworks of Scientific Discovery through Advanced Computing (SciDAC) award Fundamental Nuclear Physics at the Exascale and Beyond and the Topical Collaboration 
in Nuclear Theory 3D quark-gluon structure of hadrons: mass, spin and tomography. YZ is partially supported by the 2023 Physics Sciences and Engineering (PSE) Early Investigator Named Award program
at Argonne National Laboratory. This research used awards of computer time provided by: The INCITE program at Argonne Leadership Computing Facility, a DOE Office of Science User Facility operated under Contract DE-AC02-06CH11357,
the ALCC program at the Oak Ridge Leadership Computing Facility, which is a DOE Office of Science User Facility supported under Contract DE-AC05-00OR22725; the National Energy Research Scientific Computing Center,
a DOE Office of Science User Facility supported by the Office of Science of the U.S. Department of Energy under Contract DE-AC02-05CH11231 using NERSC award NP-ERCAP0028137. 
Part of the data analysis was carried out on Swing, a high-performance computing cluster operated by the Laboratory Computing Resource Center at Argonne National Laboratory.

\end{document}